\documentclass[reprint,showpacs,amssymb,amsmath,aps,pra]{revtex4-1}
\usepackage{graphics}
\usepackage{bm}
\usepackage{epstopdf}
\usepackage{graphicx}
\usepackage{amsmath}
\usepackage{amsthm}
\usepackage{bm}
\usepackage{bbm}
\begin{document}

\title{Separability and parity transitions in $XYZ$ spin systems under nonuniform fields} 
\author{N.\ Canosa$^1$, R.\ Mothe$^{2,1}$,  R.\ Rossignoli$^{1,3}$}
\affiliation{$^1$ Instituto de F\'{\i}sica de La Plata, CONICET, and Departamento\ de F\'{\i}sica, Universidad Nacional de La Plata,
C.C. 67, La Plata (1900), Argentina\\
$^2$ D\'epartement de Physique de l'\'Ecole Normale Sup\'erieure de Lyon, Universit\'e Claude Bernard Lyon 1, 46, all\'ee d'Italie, 69007 Lyon, France\\
$^{3}$ Comisi\'on de Investigaciones Cient\'{\i}ficas (CIC), La Plata (1900), Argentina}
\begin{abstract} 
We examine the existence of completely separable ground states (GS) in 
finite spin-$s$ arrays with anisotropic $XYZ$ couplings, immersed in a non-uniform magnetic field along one of the principal axes. The general conditions for their existence are determined. The analytic expressions for the separability curve in field space, and for the ensuing factorized state and GS energy,  are then derived for alternating solutions, valid for any spin and size. They generalize results for uniform fields and show that nonuniform fields 
can induce GS factorization also in systems which do not exhibit this phenomenon in a uniform field.  It is also shown that such curve corresponds to a 
fundamental $S_z$-parity transition of the GS, present for any spin and size, 
and that two different types of GS parity diagrams can emerge, according to the anisotropy.   
The role of factorization in the magnetization and entanglement of these systems  is also analyzed, and analytic expressions  at the borders of the factorizing curve are provided. Illustrative examples for spin pairs and chains are as well discussed.   
\end{abstract}
\maketitle
\vspace*{-1.45cm}

 \section{Introduction}
Interacting spin chains and arrays constitute paradigmatic many-body quantum systems characterized by strong quantum correlations. They conform an ideal scenario for probing and analyzing 
 entanglement,  critical behavior and other nontrivial cooperative phenomena \cite{Sa.99,ON.02,A.08}. Finite spin chains  have also been proposed as good candidates for performing different quantum processing tasks \cite{BG.07}. Interest on these systems  has been recently stimulated by the advances in quantum control techniques \cite{Ya.16,L.12}, which  make it possible to simulate finite quantum spin systems with tunable couplings and magnetic fields through different platforms \cite{N.14}, including  cold atoms in optical lattices \cite{S.11,N.14,L.12,Guy.18}, trapped ions \cite{N.14,PC.04,KK.09, B.12,A.16} and  superconducting Josephson junctions \cite{Sa.15,W.17,Ba.13}. 

 The GS of these systems is normally entangled, even if immersed in a finite  external magnetic field.  However, due to the competition between spin interactions and the external magnetic field,  the GS may become   {\it exactly} separable, i.e.\ a product of single site states, under certain conditions.  This remarkable phenomenon  occurs  at particular finite values and orientations of the magnetic field, denoted as {\it factorizing fields}. 
  It  was first  analyzed in detail in  \cite{Ku.82} for  spin chains with antiferromagnetic first neighbor $XYZ$ couplings under a uniform field, and since then it has been studied in different  spin models, mostly under  uniform magnetic fields
\cite{MS.85,T.04,DK.04,Am.06,FF.07, GI.07,RCM.08,GI.08,G.09,CRM.10,CRC.10,CRC.10,RLA.10,TR.11,CRG.13,II.13,kar.14,
CRC.15,Yi.19,TC.16,irons.17}, with a general treatment  provided in \cite{GI.08} and \cite{CRC.15}. A remarkable aspect of GS factorization is that it corresponds to a GS {\it entanglement transition}, 
in which entanglement changes its type \cite{Am.06} and, moreover, reaches {\it full range} in its immediate vicinity\cite{T.04,Am.06,RCM.08,CRM.10,CRC.15}. The critical properties of entanglement and  quantum correlations in connection with factorization  
have consequently aroused great attention \cite{T.04,Am.06,FF.07,RCM.08,CRM.10,CRC.10,TR.11,CRG.13,II.13,kar.14,CRC.15,TC.16,
irons.17,Yi.19}.

The case of non transverse factorizing fields in systems with $XYZ$ Heisenberg couplings was discussed in \cite{Ku.82} and \cite{CRC.15}, while alternating  transverse factorizing fields in  $XY$ systems ($J_z=0$) were explicitly considered in \cite{CRM.10} and \cite{TC.16}. 
On the other hand, GS separability  in chains and arrays with $XXZ$ couplings under a non-uniform field along the $z$ axis was recently examined in \cite{CRC.17}. In addition to the fully aligned phases, an exceptional multicritical factorization point where all magnetization plateaus merge was shown to exist \cite{CRC.17}, for a wide range of nonuniform factorizing field configurations and any spin and size, at which a continuous set of symmetry-breaking factorized GS's exists. Moreover, under non uniform fields $XXZ$ systems may exhibit novel and nontrivial magnetization diagrams and critical behavior \cite{CRCL.19}. 

Motivated by these results our aim here is to examine GS  factorization in finite anisotropic $XYZ$ systems of arbitrary spin under a non uniform field along one of the principal axes ($z$). 
 In contrast with the $XXZ$ case in a similar field, the eigenstates of an  $XYZ$ system no longer possess a definite magnetization along $z$, but still have a definite $S_z$-parity $P_z\propto e^{-i\pi S_z}$. And the GS magnetization transitions of the finite $XX$  \cite{CR.07,plas.09} and $XXZ$ cases \cite{CRC.17,CRCL.19} arising for increasing fields become replaced in a finite $XYZ$ array by parity transitions \cite{RCM.08,G.09,CRC.10,CRM.10}, emerging  due to the crossing of the lowest energy levels of opposite parity.  Factorization in anisotropic  $XYZ$ systems  requires, as will be seen, the breaking of the fundamental parity symmetry, entailing that it  must necessarily occur at one of these  transitions. 

After deriving the general equations for factorization  in XYZ  systems, we will show that under nonuniform fields, GS factorization will take place at a certain {\it curve} in field space, determined analytically for alternating solutions, which  represents the fields where a fundamental GS parity transition (energy level crossing), occurring for {\it any spin and finite size}, takes place.  At this curve a pair of fully separable parity breaking GSs become exactly feasible. 
Special entanglement properties will hold in its immediate vicinity for small systems, determined by the corresponding parity restored states. The GS parity diagram  will also exhibit other  parity transitions, reflecting a cascade of level crossings between the lowest states of opposite parity, whose number depends on the  total spin, i.e.\ on the system spin and size. Moreover, we will show that two distinct types of GS phase parity diagrams  can be  identified according to the  anisotropy of the coupling, separated by the critical $XZZ$ case ($J_y=J_z<J_x$)  where  all parity transition curves, including the fundamental factorizing curve, merge at zero field.

 These results also show that nonuniform fields can induce GS factorization and parity transitions for {\it any} value of the couplings, including systems which do not present  factorization or parity transitions under a uniform field.  And since general spin-$s$ anisotropic $XYZ$ systems in nonuniform fields are not exactly solvable  nor integrable \cite{Sa.99,Ta.99,Wa.17,fra.17} (with the exception of some particular spin $1/2$ cases \cite{Cl.19})  the knowledge  of  curves in field or parameter space where  exact analytic results can be obtained for any spin and size is of most importance. An interesting related aspect is that in small anisotropic $XYZ$ arrays the factorizing field can be signaled by a finite magnetization jump  \cite{RCM.08,CRM.10}. 
The feasibility of an experimental detection of  this entanglement transition in finite clustered quantum magnets  through  changes in the magnetization and neutron scattering cross section  was recently analyzed in  \cite{irons.17} for a uniform field. 

The formalism is presented in section  \ref{II}, where  analytic results for the factorizing curve and GS are derived, first for spin pairs and then for spin chains and arrays. GS parity diagrams are also discussed. Illustrative results for the GS magnetization and entanglement are provided in  \ref{III} for spin pairs and chains, in order to disclose the different role played by  factorization in these systems. 
Analytic results at the border of factorization are also determined.  
Conclusions are finally given in \ref{IV}. 

\section{Separability in $XYZ$ systems of general spin  in  nonuniform fields\label{II}}
\subsection{General separability equations}
We consider an array of $n$ spins  $s_i$  interacting through anisotropic $XYZ$ Heisenberg couplings in the presence of a nonuniform external   magnetic field along the $z$ axis.  The Hamiltonian reads 
\begin{equation}
H=-\sum_{i,\mu} h^i S^z_i - \sum_{i<j} (J_x^{ij} S^x_i S^x_j+J_y^{ij} S^y_i S^y_j+J_z^{ij} S^z_i S^z_j) \,,
\label{H1}
\end{equation}
where  $h^i$ and $S_i^\mu$, $\mu=x,y,z$, denote the field and spin components at site $i$ and  $J_\mu^{ij}=J_\mu^{ji}$ are the coupling strengths.  $H$ commutes with the global $S^z$ parity 
\begin{equation}P_z=\exp[\imath\pi\sum_i(S^z_i-s_i)]\,,\end{equation} 
 for any value of the fields or couplings, as $P_z$ just changes the sign of all $S^x_i$ and $S^y_i$. Any nondegenerate eigenstate will then have a definite parity $P_z=\pm 1$. 

We now examine the possibility of a  {\it  fully separable} exact GS $|\Theta\rangle$  of $H$, of the form
\begin{equation}
|\Theta\rangle=\otimes_{i=1}^n (R_i|\!\uparrow_i\rangle)=|\theta_1\phi_1,\theta_2\phi_2,\ldots\rangle\,,\label{Theta}
\end{equation}
where $|\!\uparrow_i\rangle$ is  the state with maximum spin along $z$ 
at site $i$ 
 ($S^z_i|\!\uparrow_i\rangle=s_i|\!\uparrow_i\rangle$) and $R_i=e^{-\imath\phi_i S^z_i}e^{-\imath\theta_iS^y_i}$ rotates it  to direction  
$\bm{n_i}=(\sin\theta_i\cos\phi_i,\sin\theta_i\sin\phi_i,\cos\theta_i)$, such that $\langle \Theta|\bm{S}_i|\Theta\rangle=s_i\bm{n}_i$. 
This state  will break parity symmetry unless  $\sin\theta_i=0 \,  \forall \,i$, and can then be an exact GS only at fields where the GS becomes  degenerate, i.e.\ {\it where a  GS parity transition takes place}.  
 
The eigenvalue equation $H|\Theta\rangle=E_\Theta|\Theta\rangle$ can be rewritten as $(\otimes_{i=1}^n R^\dagger_i)H(\otimes_{i=1}^n R_i)|0\rangle =E_\Theta|0\rangle$, 
with $|0\rangle=\otimes_{i=1}^n|\!\!\uparrow_i\rangle$ the state with all spins aligned along $z$.  This implies replacing all  $S_i^\mu$ in $H$ by the rotated operators ${S'_i}^\mu=R_i^\dagger S_i^\mu R_i$. We then 
 obtain two sets of equations, which together constitute the necessary and sufficient conditions  ensuring that  $|\Theta\rangle$ is an exact eigenstate  \cite{CRC.15}. The first set comprises the field independent equations 
\begin{subequations}\begin{eqnarray}
&&
J_y^{ij}(\cos\phi_i \cos\phi_j -\cos\theta_i\sin\phi_i\cos\theta_j\sin\phi_j)
\nonumber\\ && =J_x^{ij}(\cos\theta_i \cos\phi_i \cos\theta_j \cos\phi_j-\sin\phi_i\sin\phi_j)\nonumber\\ 
&& 
\;\;\;\;+ J_z^{ij}\sin\theta_i\sin\theta_j, \label{eq1}\\ &&
J_y^{ij}(\cos\theta_i\sin\phi_i\cos\phi_j +\cos\phi_i\cos\theta_j\sin\phi_j)
\nonumber\\
&&=J_x^{ij}(\cos\theta_i \cos\phi_i \sin\phi_j + \sin\phi_i \cos\theta_j \cos\phi_j)\,,\;\;\;\label{eq2}
\end{eqnarray}
\label{eq12}
\end{subequations}
to be satisfied for all coupled pairs $i,j$, which are also spin independent and cancel all elements of $H$ connecting $|\Theta\rangle$ with two-spin excitations (terms $\propto{S'_i}^-{S'_j}^-$). 
The second set contains the 
field dependent equations, which in the absence of field components along  $x$ and $y$  become 
\begin{subequations}
\begin{eqnarray}
h^i  \sin\theta_i&=&{\textstyle\sum\limits_{j\neq i}}s_j[\cos\theta_i  \sin\theta_j (J_x^{ij} \cos\phi_i\cos\phi_j\nonumber\\
&&+J_y^{ij}\sin\phi_i \sin\phi_j)
-J_z^{ij} \sin\theta_i \cos\theta_j]\,,
\label{eq3}\\
0 &=&{\textstyle\sum\limits_{j\neq i}}s_j  \sin\theta_j[J_x^{ij}\sin\phi_i \cos\phi_j-J_y^{ij}\cos\phi_i \sin\phi_j]\,,\;\;\;\;\;\;\;\label{eq4}
\end{eqnarray}\label{eq34}\end{subequations}
and determine the  factorizing fields $h^i$. They cancel all elements connecting $|\Theta\rangle$ with single spin excitations 
and coincide with the mean field equations 
$\partial\langle H\rangle_{\Theta}/\partial\theta_i=0$, $\partial\langle H\rangle_{\Theta}/\partial\phi_i=0$, where $\langle H\rangle_{\Theta}=\langle \Theta|H|\Theta\rangle$. 

With the replacements 
$h^i=s{h'}^i/s_i$, $J_\mu^{ij}=sj_{\mu}^{ij}/(s_i s_j)$, where $s>0$ (in principle arbitrary) can represent an average spin,  Eqs.\ (\ref{eq34}) also become {\it spin-independent} at fixed values of ${h'}^i$ and $j_\mu^{ij}$. Therefore, the present factorization is essentially a {\it spin independent phenomenon}: If present, for instance, in a spin $1/2$ array with couplings $J_\mu^{ij}=2j_\mu^{ij}$ at factorizing fields $h^i$, it will also arise in a spin-$s$ array with couplings $J_\mu^{ij}=j_\mu^{ij}/s$ at the same factorizing fields  $h^i$ (or equivalently, at rescaled fields $(2s)h^i$ if couplings $J_\mu^{ij}$ remain unaltered). The angles $\theta_i,\phi_i$ of the factorized eigenstate will remain unchanged. 
In this sense it is universal. 
In what follows we then consider a common spin $s_i=s$ $\forall$ $i$ and set 
\begin{equation}
    J^{ij}_\mu=j^{ij}_\mu/s\,,\label{jmu}
\end{equation}
such that  Eqs.\  (\ref{eq12})--(\ref{eq34}),
 as well as the scaled energy 
\begin{eqnarray}\langle H\rangle_\Theta/s
&=&
-\sum_i h^i n_i^z
-\sum_{i<j}\sum_\mu j_\mu^{ij} n_i^\mu n_j^\mu\,,
\end{eqnarray}
 are  $s$-independent at fixed fields $h^i$ and couplings $j^{ij}_{\mu}$. 

\subsection{The case of a spin-$s$ pair}
\subsubsection{General results}
We first consider a single spin-$s$ pair $i\neq j$, with $j_\mu^{ij}=j_\mu$,  $h^{i(j)}=h_{1(2)}$. We focus on the anisotropic case $j_x\neq j_y$, and choose  the $x,y$ axes such that $|j_y|< |j_x|$. We then seek solutions with $\phi_1=\phi_2=0$ and $\theta_{1(2)}\in(-\pi,\pi]$,  such that $\langle \bm{S}_i\rangle_{\Theta}$ lies in the $x,z$ plane and  $|\Theta\rangle=|\theta_1,\theta_2\rangle$ \cite{ft}. Eqs.\ (\ref{eq2}) and (\ref{eq4}) are then trivially satisfied whereas  (\ref{eq1}) and (\ref{eq3}) become  
\begin{subequations}\begin{eqnarray}
j_y&=&j_x\cos\theta_1\cos\theta_2+j_z\sin\theta_1\sin\theta_2\,, \label{eq11}\\h_1  \sin\theta_1&=&j_x\cos\theta_1 \sin\theta_2 -j_z \sin\theta_1 \cos\theta_2\,,
\label{eq22}\\
h_2  \sin\theta_2&=&j_x\cos\theta_2 \sin\theta_1 -j_z\sin\theta_2 \cos\theta_1\,.
\label{eq33}
\end{eqnarray}
\label{eqq}
\end{subequations}

\begin{figure}[ht!]
	\centerline{\scalebox{.15}{\includegraphics
{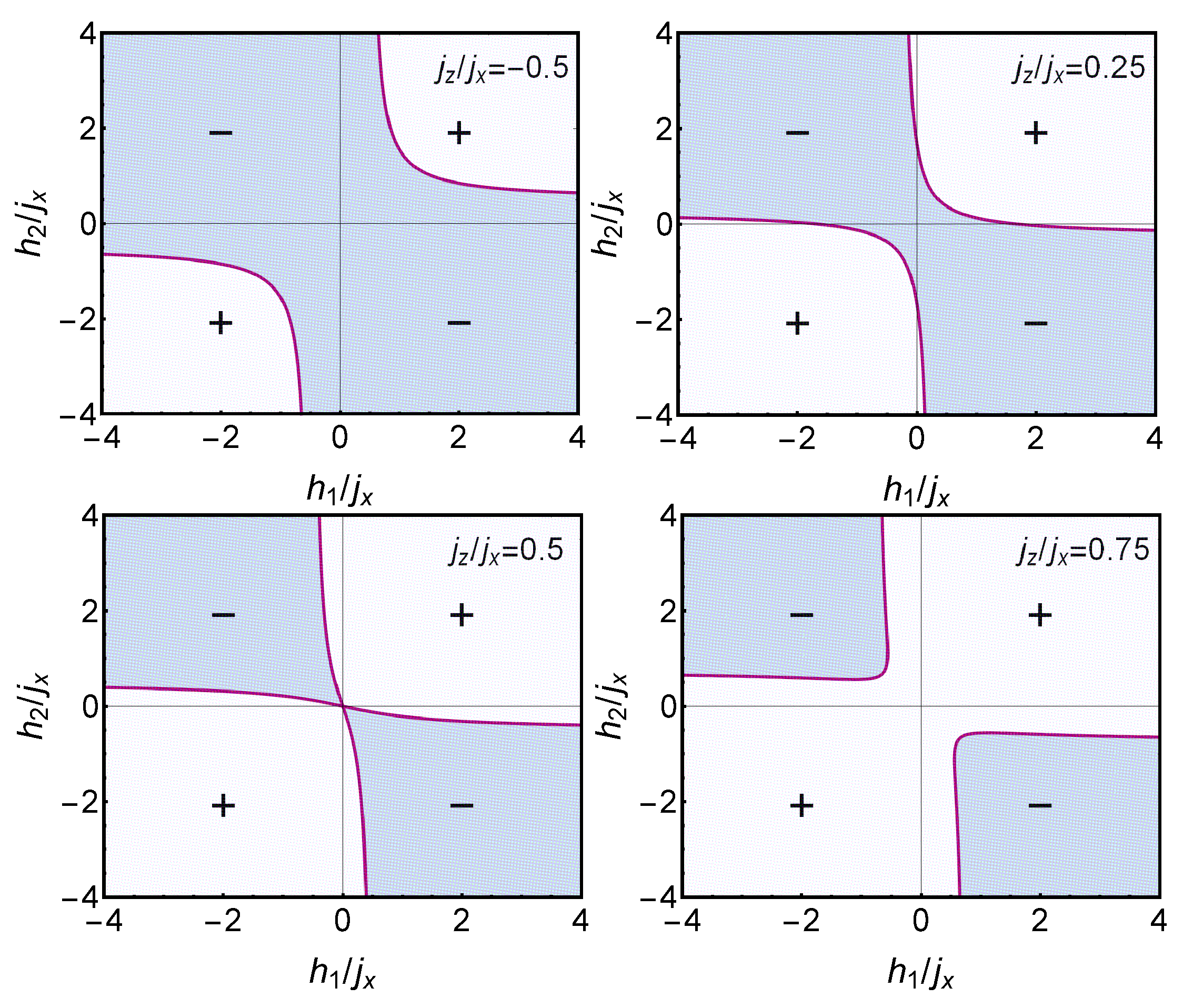}}}
	\caption{The ground state  factorization curves 
		(solid lines) in the field plane $(h_1,h_2)$ determined by Eq.\ \eqref{eq}, for $j_y=j_x/2>0$  and different values of $j_z/j_x$. At these curves a GS $S_z$-parity transition takes place. For a spin $1/2$ pair, the lighter (darker) colored sectors separated by these curves  correspond to  positive (negative) GS parity $P_z=\pm 1$. The same factorization curves remain, nevertheless, valid for a general spin-$s$ pair  as well as for  a spin-$s$  chain or lattice (see Fig.\ \ref{f3})}. 
	\label{f1}
\end{figure}

It is first seen that given arbitrary angles $\theta_{1(2)}$ with $\sin\theta_{1(2)}\neq 0$ \cite{ft2}, {\it unique} values of $j_y$ and $h_{1(2)}$ always exist such that previous equations are satisfied.  Using (\ref{eqq}) it can be shown that these values satisfy the constraints 
\begin{equation}
(h_1\pm h_2)^2+(j_x\mp j_y)^2=(\varepsilon_{\Theta}\pm j_z)^2\,,\label{aux}\end{equation}
with $\varepsilon_{\Theta}=E_\Theta/s$ the scaled pair energy at factorization: 
\begin{subequations}\label{ee12}\begin{eqnarray}\varepsilon_{\Theta}&=&
-\!\sum_{i=1,2}h_i\cos\theta_i
-j_x\sin\theta_1
\sin\theta_2
-j_z\cos\theta_1\cos\theta_2\hspace*{.9cm}\label{ee1}\\
&=&-j_x\tfrac{\sin^2\theta_1+\sin^2\theta_2-\sin^2\theta_1\sin^2\theta_2}{\sin\theta_1\sin\theta_2}+j_z\cos\theta_1\cos\theta_2\label{ee2}\,.\end{eqnarray}
\end{subequations}
For angles $\theta_{1(2)}$ such that $\varepsilon_{\Theta}\leq -|j_z|$,  Eq.\ (\ref{aux}) implies the following  constraint on the fields and couplings: 
\begin{widetext}
\begin{eqnarray} 
&&\sqrt{(h_1-h_2)^2+(j_x+j_y)^2}
-\sqrt{(h_1+h_2)^2+(j_x-j_y)^2}=2 j_z,\nonumber\\
&&\label{eq}
\end{eqnarray}
\end{widetext}
which is the fundamental factorization condition for the GS, as shown  below  and in the Appendix. For fixed couplings it determines the set of GS factorizing fields $(h_1,h_2)$, i.e.\ the {\it GS factorization curves},  
depicted in Fig.\ \ref{f1}. They  represent the fields where a  fundamental GS parity transition, arising for {\it any} spin $s$, takes place.  

It should be noticed that real fields satisfying  Eq.\ \eqref{eq}  exist for {\it any} value of the couplings $(j_x,j_y,j_z)$. 
In contrast, for a uniform field  $h_1=h_2=h$, Eq.\  \eqref{eq} can be satisfied for real $h$ only if $j_z\leq j_y\, {\rm Sgn}(j_x)$ (for $|j_y|< |j_x|$), in agreement with \cite{T.04,RCM.08}. Thus,  nonuniform fields can induce  GS factorization, and hence a GS parity transition, {\it for  any value of the couplings},  including ferromagnetic ($j_\mu>0$) and antiferromagnetic ($j_\mu<0$) cases. 

Eq.\ (\ref{aux}) also enables the following closed analytic expressions for the pair energy  (\ref{ee12}) at factorization: 
\begin{subequations}\begin{eqnarray}
\varepsilon_\Theta&=&-\tfrac{\sqrt{(h_1-h_2)^2+(j_x+j_y)^2}+\sqrt{(h_1+h_2)^2+(j_x-j_y)^2}}{2}\;\;\;\label{ee22}\\ 
&=&-\sqrt{h_1^2+h_2^2+j_x^2+j_y^2-j_z^2}\label{ee3}\\
&=&-(j_x j_y-h_1 h_2)/j_z\label{ee4}\,,
\end{eqnarray}
\label{ee34}
\end{subequations}
where Eq.\ (\ref{eq}) is assumed to be satisfied 
and \eqref{ee4} holds for $j_z\neq 0$ (for $j_z\rightarrow 0$, $h_1h_2\rightarrow j_xj_y$, see Eq.\ \eqref{eqxxz}). It is then verified from \eqref{eq}-\eqref{ee22}  that $\varepsilon_{\Theta}\leq-|j_z|$. Angles implying other signs of $\varepsilon_{\Theta}\pm j_z$ lead to different
signs of the square roots in \eqref{eq} and \eqref{ee22}--\eqref{ee3},  and correspond
to crossings of excited states of opposite parity, i.e., to
factorization of excited states.

It is also possible to obtain from 
(\ref{eqq}) and \eqref{eq} an  analytic expression for  $\cos^2\theta_{i}$ 
in terms of  its own field $h_{i}$:  
\begin{equation}
\cos^2\theta_{i}=
\frac{h_i^2+ j_y^2-j_z^2}{h_i^2+j_x^2-j_z^2}\,,
\;\;\;\;i=1,2\,,\label{thij}
\end{equation}
where  Eq.\ \eqref{eq} is again assumed to be fulfilled. 
The sign of $\cos\theta_i$ should be such that  Eqs.\ (\ref{eqq}) are satisfied. 
 In the uniform case $h_1=h_2$, Eqs.\ \eqref{eq} and \eqref{thij} imply, for $j_z\leq \gamma j_y$ ($\gamma={\rm Sgn}(j_x)$), the known results $|h_i|=
 \sqrt{(j_x-\gamma j_z)(j_y-\gamma j_z)}$,  $\cos^2\theta_i=\frac{j_y-\gamma j_z}{j_x-\gamma j_z}$ \cite{MS.85,T.04,Am.06,FF.07,RCM.08,GI.08,CRM.10} (which implies the result $|h_i|=
 \sqrt{j_x j_y}$ for the $XY$ case  \cite{McK.11}).  

We notice that  $(\theta_1,\theta_2)$ and $(-\theta_1,-\theta_2)$ are degenerate solutions: they lead in (\ref{eqq}) to the same $j_y$, fields $h_{1(2)}$ and energy $\varepsilon_\Theta$, in agreement with parity symmetry ($P_z|\Theta\rangle=|-\Theta\rangle\equiv|-\theta_1,-\theta_2\rangle$), showing explicitly  the degeneracy at  factorization.
We also notice that  $(\theta_1,\theta_2)$ and ($\pi-\theta_1,\pi-\theta_2)$  lead to the same $j_y$ but opposite fields,  in agreement with a global $\pi$ rotation around the $x$ axis,  whereas $(\theta_1,\theta_2)$ and ($\theta_1,-\theta_2)$ correspond to couplings $(j_x,j_y,j_z)$ and $(-j_x,-j_y,j_z)$ respectively, with the same fields,  in agreement with a $\pi$ rotation around the $z$ axis of the second spin. Hence, these cases have all the same spectrum and GS energy, as verified in  Eqs.\ \eqref{ee34}. For $j_x>0$, minimum energy requires 
$\theta_1$ and $\theta_2$ of the same sign (for $\theta_i\in(-\pi,\pi]$), as seen from  \eqref{ee1}. 

\subsubsection{The spin $1/2$ case \label{B2}}
It can be easily verified  that for a spin $1/2$ pair, Eq.\ (\ref{eq}) determines precisely the fields where the {\it GS  parity transition} takes place: The exact energies of the spin $1/2$ pair, obtained from diagonalization, are 
\begin{subequations}
\label{Epm}
\begin{eqnarray}
E^+_{\pm}&=&\tfrac{1}{2}[\pm\sqrt{(h_1+h_2)^2+(j_x-j_y)^2}-j_z]\label{e1}\,,\\
E^-_{\pm}&=&\tfrac{1}{2}[\pm\sqrt{(h_1-h_2)^2+(j_x+j_y)^2}+j_z]\label{e2}\,,
\end{eqnarray}
\end{subequations}
 where $E^+_{\pm}$ ($E^{-}_{\pm}$) correspond to the positive (negative) parity eigenstates  
\begin{subequations}

\label{psipm}\begin{eqnarray}
|\Psi^{+}_{\pm}\rangle&=&\cos\gamma^+_{\pm}|\uparrow\uparrow\rangle+\sin\gamma^+_{\pm}|\downarrow\downarrow\rangle\,,\label{psi11}\\
|\Psi^{-}_{\pm}\rangle&=&\cos\gamma^-_{\pm}|\uparrow\downarrow\rangle+\sin\gamma^-_{\pm}|\downarrow\uparrow\rangle\,,
\label{psi22}\end{eqnarray}\end{subequations}
with
$\tan\gamma^{\pm}_\nu=-\frac{\nu\sqrt{(h_1\pm h_2)^2+(j_x\mp j_y)^2}+h_1\pm h_2}{j_x\mp j_y}$ ($\nu=\pm$). 
The lowest energies $E^{\pm}_-$ for each parity then cross when $E^+_-=E^-_-$, which leads to Eq.\ (\ref{eq}). And at the crossing, after solving for $j_z$ the scaled GS energy $E^\pm_-/s$  becomes identical with Eqs.\ (\ref{ee34}). Similarly,  crossing of excited opposite parity levels leads to different signs of the square roots in (\ref{eq}), implying $\varepsilon_{\Theta}\pm j_z$ not both negative. 

The connection between the entangled definite parity eigenstates  (\ref{psipm}) and the separable parity breaking eigenstates $|\pm\Theta\rangle=|\pm\theta_1,\pm\theta_2\rangle$, with $|\pm\theta\rangle=\cos\frac{\theta}{2}|\!\!\uparrow\rangle\pm\sin\frac{\theta}{2}|\!\!\downarrow\rangle$ for $s=1/2$, is just parity projection: 
\begin{equation}
|\Psi^{\pm}_-\rangle=\frac{|\Theta\rangle\pm |-\Theta\rangle}{\sqrt{2(1\pm\langle -\Theta|\Theta\rangle)}}\propto \frac{\mathbbm{1}\pm P_z}{2}|\Theta\rangle\,.\label{rel}
\end{equation}
 Eq.\ \eqref{rel} holds only at the GS factorization curve (\ref{eq}), where it can be verified that $\tan\gamma^{\pm}_-=\tan\frac{\theta_1}{2}\tan^{\pm 1}\frac{\theta_2}{2}$, with $\theta_{1(2)}$ obtained from \eqref{thij} and fulfilling \eqref{eqq}. The proof for  general spin $s$ is given in the Appendix.

\subsubsection{Factorization curves} 
Eq.\ (\ref{eq}) determines a pair of curves in the  field plane $(h_1,h_2)$, depicted in Fig.\ \ref{f1} for $j_x>0$.  They are valid  for any spin when couplings are scaled as in (\ref{jmu}), and are  symmetric with respect to the $h_1=h_2$ and $h_1=-h_2$ lines. For a spin $1/2$ pair they separate the positive from the negative GS parity sectors,  determining in this case  all GS parity transitions as the fields are varied. 

For $|j_y|<j_x$, and setting also $j_y\geq 0$ with no loss of generality (its sign can be changed by a $\pi$ rotation  around the $x$ axis of the second spin, such that $h_2\rightarrow -h_2$ and $j_\mu\rightarrow-j_\mu$ for $\mu=y,z$), three  distinct cases arise:\\
a) $j_z<j_y$ (upper panels): In this case the GS has  negative parity at zero field $h_{1(2)}=0$ and the factorizing curves have vertices (minimum of $h_1^2+h_2^2$) at 
\begin{equation}
h_1=h_2=\pm \sqrt{(j_x-j_z)(j_y-j_z)}\,,\label{h12}
\end{equation}
(uniform field), where $\varepsilon_{\Theta}=-(j_x+j_y-j_z)$ (Eqs.\ \eqref{ee34}) and  the factorized state is uniform ($\theta_1=\theta_2)$.   Factorization for non-uniform fields is in this case the smooth continuation of that obtained for the uniform case. \\ 
b) $j_z>j_y$ (bottom right panel): Here the GS has positive parity at zero field and vertices lie at {\it opposite} fields 
\begin{equation}h_1=-h_2=\pm\sqrt{(j_z-j_y)(j_z+j_x)}\,,\label{h12m}
\end{equation}
where $\varepsilon_{\Theta}=-(j_x+j_z-j_y)$ and $\theta_2=\pi-\theta_1$  (Eq.\  (\ref{thij})). In this case no GS factorization occurs at uniform fields $h_1=h_2$, as previously stated. Moreover, GS factorization requires always fields of {\it opposite sign}. \\
c) $j_z=j_y$ (bottom left panel): In this critical case   (transition between previous cases a and b) the  factorization curves, which already involve fields $(h_1,h_2)$ of opposite sign,  intersect at the origin $h_{1(2)}=0$,  where  all sectors meet. At this point,  $\varepsilon_{\Theta}=-j_x$ and $\theta_1=\theta_2=\pm\pi/2$, 
implying that the factorized GS's  $|\pm\Theta\rangle$ are here {\it orthogonal} for any spin $s$ (they are fully aligned states along the $\pm x$ directions). The parity restored states (\ref{rel})  then become {\it Bell-type states} at this point. 

The curves asymptotes lie at $h_{1(2)}=\pm j_z$ in all cases, since  for strong field  $h_{1(2)}\rightarrow\pm\infty$,  Eq.\ (\ref{eq}) leads to  
$h_{2(1)} \rightarrow \mp j_z$.   
For $0<j_z<j_y$ the separability curves then cross the axes (top right panel in Fig.\ \ref{f1}), implying that the factorizing field at one of the spins 
will vanish and change sign. Hence  GS factorization (and thus the GS parity transition) can in this case  be achieved {\it with just  one field}, i.e.\ with no field at one of the spins: Setting $h_{2(1)}=0$ in Eq.\ (\ref{eq}), the factorizing field at the other spin takes the value ($0<j_z<j_y$)   
\begin{equation}h_{1(2)}=\pm\sqrt{(j_x^2-j_z^2)(j_y^2-j_z^2)}/j_z\;\;\;(h_{2(1)}=0)\,.\end{equation}
 On the other hand, 
for $j_z<0<j_y$, $|h_i|\geq -j_z$, implying that finite fields of the same sign at both spins  are required for factorization in order to overcome the antiferromagnetic $j_z$ coupling (top left panel). And for $j_z>j_y$ (ferromagnetic $j_z$ coupling, bottom right panel), fields $h_1,h_2$ are again both non-zero but have opposite signs, with $|h_{i}|\geq  \sqrt{j_z^2-j_y^2}$ along the factorizing curves, as implied by \eqref{thij}. 

We also remark that Eq.\  (\ref{eq})  generalizes separability equations obtained for particular more symmetric cases, unifying them all in a single equation. For example, in the $XY$ case $j_z=0$, 
 Eq.\ (\ref{eq}) leads to a simple {\it hyperbola} in the field plane $h_1,h_2$, namely 
\begin{equation}h_1 h_2=j_x j_y\;\;\;\;(j_z=0)\,,\label{eqxxz}\end{equation} 
in agreement with results of \cite{CRM.10,TC.16}. 
And in the $XXZ$ limit $j_y\rightarrow j_x$, Eq.\ (\ref{eq}) leads to the two hyperbola branches 
\begin{equation}(h_1 \pm j_z) (h_2 \pm j_z) = j_x^2\,,\;\;(j_x=j_y)\end{equation}
where the $+$ ($-$) sign holds for  $h_1+h_2>0$ ($<0$).     They are precisely those delimiting the fully aligned phases ($S_i^z=s$ or $-s$ $\forall$ $i$)  in the $XXZ$ system \cite{CRC.17}: In this limit Eq.\ \eqref{thij} leads to $\cos^2\theta_i\rightarrow 1$, i.e., $\theta_i\rightarrow 0$ or $\pi$. Finally, if $j_z>j_y$ and $h_1=-h_2=h$, 
Eq.\ \eqref{eq} leads to
\eqref{h12m}, which in the $XXZ$ limit $j_y=j_x$ coincides with the antiparallel separability field  $h_s=\pm\sqrt{j_z^2-j_x^2}$ that 
determines  the {\it multicritical  point} in $XXZ$ systems \cite{CRC.17,CRCL.19}. Here more general solutions are feasible which break the continuous symmetry of $H$ \cite{CRC.17}. 

\begin{figure}[t]
\centerline{\scalebox{.275}{\includegraphics
{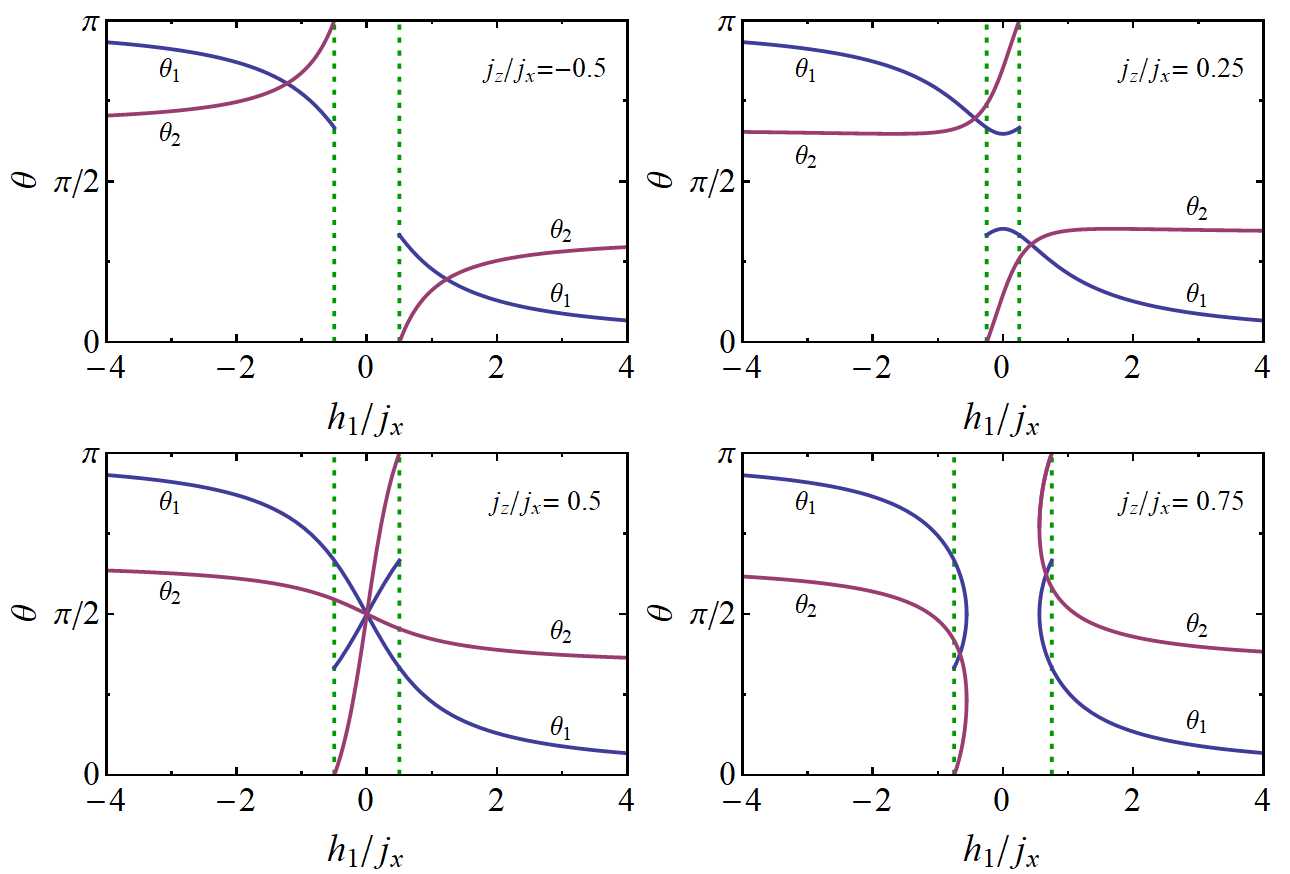}}}
\caption{The angles $\theta_{1(2)}$ 
that characterize the separable GS  along the factorization curves depicted in Fig.\ \ref{f1}, as a function of the scaled magnetic field $h_1$ (with $j_y=j_x/2>0$).  
Dashed vertical lines indicate the  asymptotes  $h_1=\mp j_z$ where $h_2\rightarrow\pm\infty$ and $\theta_2\rightarrow 0$ or $\pi$. For $j_z=j_y$ (bottom left panel) all angles approach $\pi/2$ at zero field.}
\label{f2}
\end{figure}

\subsubsection{Factorizing angles}
The positive angles $\theta_1,\theta_2$ determining  the separable GS for $j_x>0$, obtained from \eqref{thij} and satisfying \eqref{eqq}, are depicted in Fig.\ \ref{f2} as a function of the field $h_1$ (with $h_2$ obtained from \eqref{eq}), for the cases of Fig.\ \ref{f1}. The angles along  both branches of the factorizing curve are related by $\theta_i(-h_1)=\pi-\theta_i(h_1)$. In all cases, for $h_{1(2)}\rightarrow\pm\infty$ ($h_{2(1)}\rightarrow \mp j_z$),  $\theta_{1(2)}\rightarrow^{\;0}_{\;\pi}$   due to alignment with the field, as verified in all panels 
($\theta_{i}$ vanishes as $\approx\sqrt{j_x^2-j_y^2}/h_i$ for  $h_i\rightarrow\infty$). 

For  $j_z<j_y$ (top  panels) both angles $\theta_{1(2)}$ always lie and stay in the same quadrant (first or second) for each curve, coinciding at the vertex \eqref{h12}. 
In this case, $\theta_i$ is  a decreasing function of its own field strength  $|h_i|$ in the right curves, evolving for $j_z<0$ from $\theta_0$ to $0$ as $h_i$ increases from $-j_z$ to $\infty$,  with $\cos\theta_0=j_y/j_x$ (top left), 
while for  $0<j_z<j_y$, $\theta_i$ is maximum when $h_i=0$ (top right). This behavior holds up to the critical  case $j_z=j_y$ (bottom left panel), where all angles along all curves approach $\pi/2$ for  $h_{1(2)}\rightarrow0$ and hence all trajectories meet. 

In contrast, for $j_z>j_y$ (bottom right panel) the factorizing fields have opposite signs and lead to a larger difference between the angles, which now are never coincident and may lie in different quadrants.  As  $h_1$ decreases from $+\infty$ in the right branch, $\theta_1$ evolves now from $0$ to $\pi-\theta_0$, with $\cos\theta_0=j_y/j_x$,   while $\theta_2$ evolves from $\theta_0$ to $\pi$, both crossing  $\pi/2$ when the respective field becomes minimum ($|h_i|=h_{\rm min}=\sqrt{j_z^2-j_y^2}$), as seen from \eqref{thij}. The two values of $\theta_i$ along the same branch arising for $h_{\rm min}\leq h_1\leq j_z$ correspond to the two different sectors of the factorizing curve ($h_2$ above and below $-h_{\rm min} j_x/j_y$).  

\begin{figure}[ht]
\centerline{\scalebox{0.096}{\includegraphics
{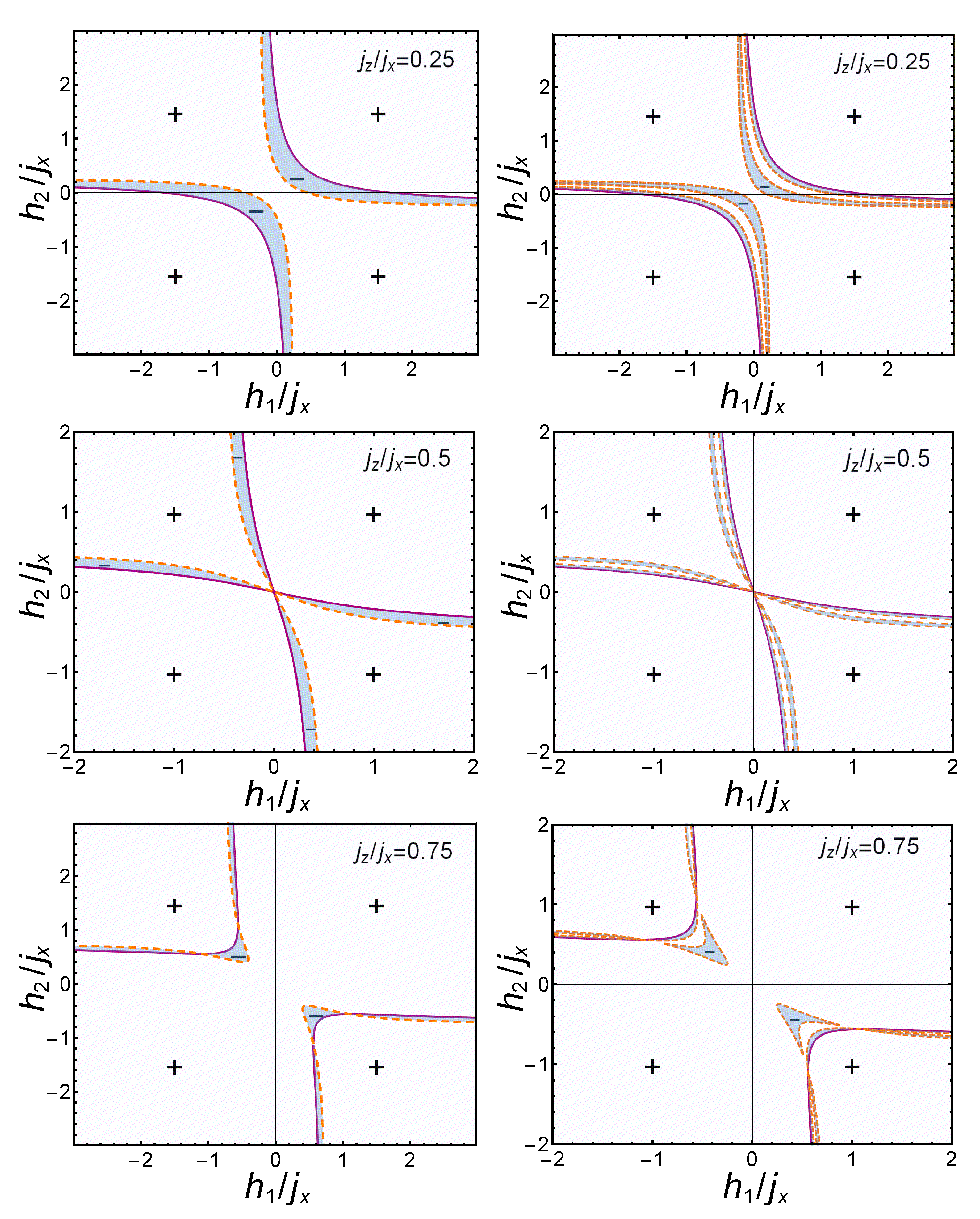}}}
\caption{GS spin parity phase diagram in the $(h_1,h_2)$ field space for a  pair of spins $1$ (left) and for a spin $1/2$ chain of  8 spins (right), for  $j_y=0.5 j_x$ and three  values of $j_z/j_x$.  Solid lines depict the factorizing curves (the same as those of Fig.\ \ref{f1})  while dashed lines the remaining GS parity transitions. Signs denote the GS parity in each sector, with dark coloured regions indicating negative parity. For $j_z=j_y$ (central panels) all curves and GS parity sectors  meet at the origin. }
\label{f3}
\end{figure}
\subsubsection{Ground state parity diagrams} 
While for a spin $1/2$ pair the  factorization curves correspond to the unique GS parity transition, 
 for higher spins $s\geq 1$ the pair  actually exhibits {\it $2s$ GS  parity transitions curves in each sector} ($4s$ curves in the whole plane) as seen on the left panels of Fig.\ \ref{f3} for $s=1$. These transitions are reminiscent of the GS  magnetization transitions of the $XX$  \cite{CR.07,plas.09} and $XXZ$ cases \cite{CRC.17,CRCL.19}. 
 
 For $j_z<j_y$ (case a) the factorization curves determine  the {\it last} GS parity transition as seen from the origin (top left panel). On the other hand, for $j_z=j_y$ (critical case c)  all GS parity transition curves intersect again at the origin (center left), where  all GS parity sectors meet. Nevertheless,  the GS remains here twofold degenerate  for any spin (we recall that at zero field,  
 $\theta_1=\theta_2=\pm\pi/2$ and  the degenerate factorized GS's are orthogonal maximally aligned states along the $\pm x$ directions).  
 
 Finally, for $j_z>j_y$ (case b)  the GS parity diagram becomes more complex (bottom panel). 
 Here the second transition curve crosses the factorization curve away from the origin, so that the narrow negative parity sector may be located ``above'' or below the latter. The energy gap between the two lowest states of opposite parity is, nevertheless, very small in the sector of negative GS parity for this anisotropy (it increases towards the $XXZ$ limit).  The behavior for higher spins is analogous  and qualitatively similar to that of a spin chain with the same total spin (right panels, see next section).  

\subsection{Extension to spin$-s$ arrays}
\subsubsection{General results} 
We now consider a general spin-$s$ array with couplings satisfying  $|j_y^{ij}|\leq j_x^{ij}$ for all interacting pairs, and analyze the possibility of a product GS  $|\Theta\rangle=|\theta_1,\theta_2,\ldots\rangle$ with $\phi_i=0$ $\forall$ $i$. Eqs.\ (\ref{eq11})--(\ref{eq33}) are to be replaced by  
\begin{eqnarray}
    j_y^{ij}&=&j_x^{ij}\cos\theta_i\cos\theta_j+j_z^{ij}\sin\theta_i\sin\theta_j\,,\label{eqgral}\\
h^i\sin\theta_i  &=& \sum_{j\neq i} j^{ij}_x\cos\theta_i\sin\theta_j - j^{ij}_z  \cos\theta_j\sin\theta_i\,,\label{hgral}
\end{eqnarray}
to be fulfilled for all interacting pairs $i\neq j$ and sites $i$. 

For $\sin\theta_i\neq 0$, Eq.\ (\ref{hgral}) can be expressed as $h^i=\sum_{j\neq i}h^{ij}$,  with $h^{ij}=j^{ij}_x\cos\theta_i\frac{\sin\theta_j}{\sin\theta_i}-j_z^{ij}\cos\theta_j$  the partial field at $i$ due to spin $j$. Thus, for arbitrary positive angles $\theta_i\in (0,\pi)$  there are always unique fields  $h^i$ and couplings $j_y^{ij}$ satisfying these equations. And  if $|j_y^{ij}|< j_x^{ij}$ for all interacting pairs, we may use the same arguments of the Appendix to show that the corresponding eigenstate  $|\Theta\rangle$, 
 together with its partner state $|-\Theta\rangle=P_z|\Theta\rangle$, will be {\it ground states}, implying 
  that this factorization will occur at a GS parity transition. The couplings $j_\mu^{ij}$ and the {\it partial} fields $h^{ij}$ will again satisfy Eq.\ (\ref{eq}) 
 for all coupled pairs $i\neq j$, i.e.,
 \begin{equation}\sum\limits_{\nu=\pm}\nu\sqrt{(h^{ij}-\nu h^{ji})^2+(j_x^{ij}+ \nu j_y^{ij})^2}=2j_z^{ij}\label{eqgral2}\,,\end{equation}
 leading in general to a nonuniform total field $h^i=\sum_{j}h^{ij}$. 
 The total GS energy at factorization will be \begin{equation}E_{\Theta}=s\sum_{i<j}
 \varepsilon_{\Theta}^{ij}\,,\label{ET}\end{equation}
 with $\varepsilon_{\Theta}^{ij}$ given by Eqs.\ \eqref{ee34} in terms of the partial fields $h^{ij}$, e.g., 
 $\varepsilon_{\Theta}^{ij}=
 -(j_x^{ij}j_y^{ij}-h^{ij}h^{ji})/j_z^{ij}$. 
 
 \begin{figure}[ht!] \centerline{\scalebox{0.085}{\includegraphics
	{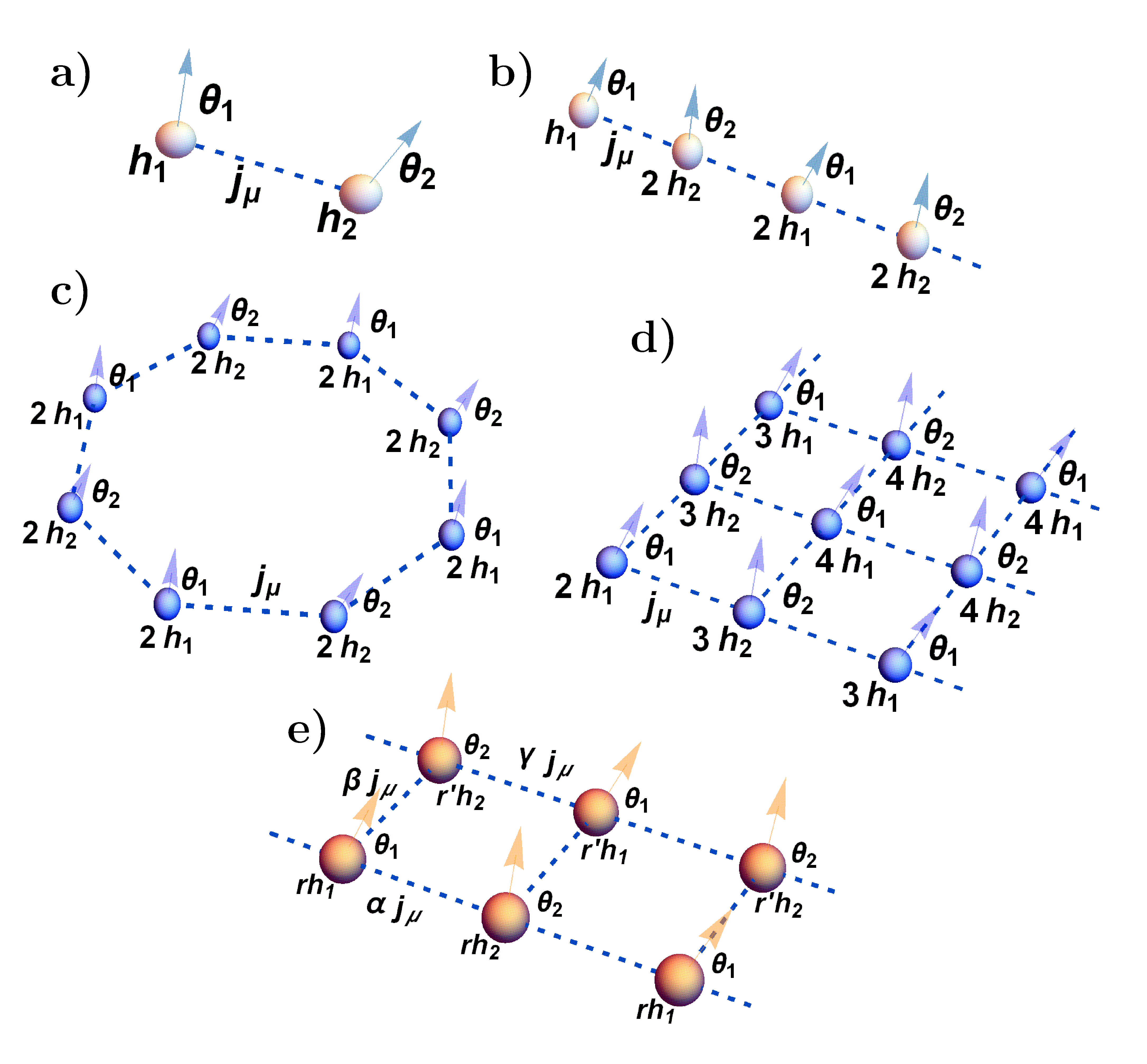}}}
\caption{Examples of spin arrays with first neighbor anisotropic $XYZ$ couplings under a non uniform field  (schematic representation), which possess an alternating separable GS for any spin $s$ when the indicated fields $h_1$ and $h_2$ satisfy Eq.\ \eqref{eq}: a) Spin pair;  b) open chain; c) cyclic chain; d) square lattice; e) ladder with non uniform couplings. 
Here the factorizing fields are $rh_{1(2)}$ ($r' h_{1(2)}$) in the lower  (upper) row, with $r=2\alpha+\beta$ 
($r'=2\gamma+\beta$).} 
\label{f4}
\end{figure}
  
\subsubsection{Alternating solutions} 
We will focus on {\it alternating} product eigenstates involving just two angles $\theta_1,\theta_2$, such that all coupled pairs are in the same product state.  These states can be exact GS's in spin chains and square-type lattices with uniform first neighbor couplings under alternating fields, 
as those depicted in Fig.\ \ref{f4}. 

We start with a one-dimensional spin chain of $n$ spins with couplings 
\begin{equation} 
j^{ij}_\mu=\delta_{i,j\pm 1}j_\mu \label{rj}\,.
\end{equation}
It is apparent that the previous product GS 
$|\Theta\rangle=|\theta_1,\theta_2\rangle$ for a single pair 
turns into an {\it alternating}  product GS  $|\Theta\rangle=|\theta_1,\theta_2,\theta_1,\ldots\rangle$ for the whole chain under an alternating field (Fig.\ \ref{f4}, b and c).   Eq.\  (\ref{eqgral}) then reduces to  (\ref{eq11}) for all coupled pairs  $(i,i\pm 1)$, while 
 (\ref{hgral}) leads to 
\begin{eqnarray}
h^i\sin\theta_i&=& r_i(j_x\cos\theta_i\sin\theta_j - j_z \sin\theta_i \cos\theta_j)\,,\label{eq222}
\end{eqnarray}
where for $i$ odd (even), $\theta_i=\theta_{1(2)}$ while $\theta_j=\theta_{2(1)}$, 
and $r_i$ is the number of spins coupled to spin $i$  ({\it coordination number}). Eq.\ (\ref{eq222}) is thus equivalent to Eqs.\ (\ref{eq22})--(\ref{eq33}) except for the factor $r_i$, 
which implies  a rescaling of the factorizing fields $h^i$: 
\begin{equation} h^i=r_i h_{1(2)}\,,\label{hi}\end{equation}
for $i$ odd (even), where $h_{1(2)}$ are the single  pair fields satisfying Eq.\ (\ref{eq}). 

In  a cyclic chain ($n+1\equiv 1$, $n$ even)  
$r_i=2$ for all spins, implying alternating factorizing fields $(2h_1,2h_2,2h_1\ldots)$ (plot c).   The same holds  in an open chain for inner spins, while for edge spins ($i=1$ or $n$) $r_i=1$, implying factorizing fields $(h_1,2h_2,2h_1,\ldots)$ (plot b). Thus, alternating product GS's are  exactly feasible in  {\it both cyclic and  open chains under alternating fields}, provided border field corrections are applied in the open case. 

These arguments also hold for $2d$ square lattices (plot d) in Fig.\ \ref{f4})   
as well as $3d$ cubic lattices with first neighbor uniform couplings, again of any size.  In these cases a similar alternating product GS $|\Theta\rangle$ remains  exactly feasible since  Eq.\ (\ref{eqgral}) reduces to Eq. (\ref{eq11}) for all coupled pairs.  The coordination number in the square lattice is $r_i=4$ for bulk spins and  $r_i=3$ ($2$) for edge (corner) spins (plot d)  while in the cubic lattice  $r_i=6$ for bulks spins and $r_i=5,4,3$ for side, edge and  corner spins respectively. In these cases  $\theta_{1(2)}$ 
are  the angles 
at sites $(i,j)$ with $i+j$ even (odd) in the square lattice ($i,j=1,2,\ldots$),  and sites $(i,j,k)$ with $i+j+k$ odd (even) in the cubic lattice.

  Thus, if $h_{1(2)}$ denote the fields satisfying the original 
 pair factorization equation (\ref{eq}), such that the angles 
 $\theta_{1(2)}$ can be obtained from Eq.\  (\ref{thij}), the factorizing fields $h^i$ for alternating product states in such arrays will be $r_i h_{1(2)}$. And the exact GS  energy (\ref{ET}) along the factorization curves (\ref{eq}) becomes just  
\begin{equation} 
E_{\Theta}=N
s\,\varepsilon_{\Theta}\label{El}\,,\;\;N=\frac{1}{2}\sum_i r_i\,,
\end{equation}
where  $\varepsilon_{\Theta}$ is the 
 pair energy (\ref{ee34}) and $N$ 
 is the total number of coupling links. For instance, in a $1d$  cyclic array of $n$ spins ($n$ even), $r_i=2$ $\forall$ $i$ and $N=n$, whereas in an open chain of $n$ spins ($n$ arbitrary) $N=n-1$. On the other hand, 
in a finite open $2d$ square lattice of $n=m\times l$ spins, $N=2n-m-l$, while in open $3d$ cubic arrays of $n=m\times l\times k$ spins, $N=3n-ml-mk-lk$. 

Previous alternating product GS's  remain  also valid  for arrays with nonuniform first neighbor $XYZ$ couplings with fixed anisotropy ratios, i.e., 
\begin{equation}
J_\mu^{ij}=r_{ij}j_\mu\,,\;\;\mu=x,y,z\,,
\end{equation}
for first neighbors $i,j$, since Eq.\ (\ref{eqgral}) still reduces to (\ref{eq11}) for all coupled pairs. Assuming $r_{ij}>0$ and $|j_y|< |j_x|$,  the final effect is again just a factor $r_i=\sum_{j} r_{ij}$ in the factorizing fields $h^i$ (Eq.\ \eqref{hi}),  as \eqref{hgral} reduces to \eqref{eq222} at all sites. This enables, for instance, direction  dependent couplings in square-type arrays and lattices (panel e) in Fig.\ \ref{f4}).  The total GS energy will still be given by Eq.\ (\ref{El}) with the present values of $r_i$. We finally remark that present results are valid for arbitrary values of the couplings $j_\mu$. The case with $j_x<0$ can be again converted to $j_x>0$ by a $\pi$ rotation around the $z$ axis at odd sites.  

\subsubsection{Ground state parity diagrams} 
The exact GS parity diagram of a spin-s chain of $n$ spins   exhibits  $2\times ns$ parity transition curves in the whole field plane $h_1,h_2$, as seen in the right panels of Fig.\ \ref{f3}, resembling those of a spin pair with the same total spin. For $j_z< j_y$ the factorization curve represents  the last GS parity transition as seen from the origin $h_1=h_2=0$,  with the GS 
reaching the final $P_z=+1$ phase beyond this curve (top right panel in Fig.\ \ref{f3}). This behavior holds up to the limit case $j_z=j_y$, where  all curves, and hence all GS parity sectors, coalesce at the origin (central right panel), and already involve fields $(h_1,h_2)$ of opposite sign.  

The diagram becomes again more complex for $j_z>j_y$ (bottom right panel), where all parity transition curves take place at  fields $h_1, h_2$ of opposite sign. The trend seen for the spin $1$ pair becomes  more notorious, with  other parity transition curves crossing the factorization curve. Hence, the factorization curve in finite XYZ chains determines the onset or border  of the region where a cascade of GS transitions between the two lowest, closely lying states of opposite parity, takes place.  It should be noticed that the energy splitting between these states in the narrow regions between curves is small and rapidly decreases with increasing size. 
In the thermodynamic limit they become degenerate and the series of parity transitions approach a continuum, with the factorization curve lying normally before the thermodynamic GS transition \cite{T.04,Am.06,FF.07,RCM.08}.

Nevertheless, factorization offers the possibility to ``extract'', along the factorizing curve, a separable nondegenerate GS just by applying an additional non uniform field $\bm{h}^i_a$ parallel at each site to the spin alignment direction $\bm{n}_i$  of the factorized GS. This field will remove the GS degeneracy and lower the chosen product state energy by an amount $-s\sum_i |h^i_a|$ (it will remain an exact GS for any strength  $|h^i_a|$),  enabling an arbitrarily large gap with the first excited state.

\section{Entanglement and  magnetization\label{III}}
\subsection{Expressions at factorization}
As the factorization curve is approached in the field plane $(h_1,h_2)$, the side-limits of physical observables and  entanglement measures will be determined by the parity restored GS's of the form (\ref{rel}),    $|\Psi^{\pm}\rangle=\frac{|\Theta\rangle\pm|-\Theta\rangle}{\sqrt{2(1\pm\langle -\Theta|\Theta\rangle)}}$, 
since the exact GS possesses  definite parity in the immediate vicinity. These states are entangled and lead to critical entanglement properties 
\cite{RCM.08,CRM.10,CRC.15}, observable  in sufficiently small systems. 

 The reduced state of a single spin $i$ in  the states $|\Psi^{\pm}\rangle$ 
 is given by 
 \begin{equation}
     \rho_i^{\pm}=\frac{|\theta_i\rangle \langle\theta_i|+|-\theta_i\rangle\langle-\theta_i|\pm\gamma_i(|\theta_i\rangle\langle-\theta_i|+|-\theta_i\rangle\langle\theta_i|)}{2(1\pm\langle-\Theta|\Theta\rangle)}\label{rhoi}
 \end{equation}
 where $\langle-\Theta|\Theta\rangle=
\prod_i\cos^{2s}\theta_i$ is the overlap of the two factorized GS's and 
$\gamma_i=\prod_{j\neq i}\cos^{2s}\theta_j=\langle-\Theta|\Theta\rangle/\cos^{2s}\theta_i$ is the complementary overlap. Thus, for any $s$, 
 $\rho_i^{\pm}$ is always a rank $2$ mixed state with two non-zero eigenvalues 
 \begin{equation}p^{\pm}_{\nu}=\frac{(1+\nu\cos^{2s}\theta_i)(1\pm\nu\gamma_i)}
 {2(1\pm\langle-\Theta|\Theta\rangle)}\,,\;\;\;\nu=\pm 1\,,\label{ppm}\end{equation}
 with  $p^\pm_++p^\pm_-=1$. 
 The ensuing single spin magnetization $\langle \bm{S}_i\rangle={\rm Tr}\,\rho_i^{\pm}\bm{S}_i$, which in a definite parity state always points along  $z$     
 (${\rm Tr}\rho_i^{\pm} S_i^{\mu}=0$ for $\mu=x,y$), is 
 \begin{equation}
 \langle S_i^z\rangle_{\pm}={\rm Tr}\,\rho_i^{\pm}\,S_i^z=s\frac{\cos\theta_i(1\pm\gamma_i
 \cos^{2s-2}\theta_i)}{1\pm\langle-\Theta|\Theta\rangle}\label{Sz}\,.
 \end{equation}
 This  leads to a {\it magnetization step} at the parity transition \cite{RCM.08, CRM.10}, visible for small sizes and spin. If $\langle-\Theta|\Theta\rangle$ and $\gamma_i$ are neglected, 
 we obviously obtain $\langle S_i^z\rangle_{\pm}\approx s\cos\theta_i$. 
 
 The entanglement of spin $i$ with the rest of the chain can be conveniently  measured 
 through the linear entropy $S_2(\rho_i)=2(1-{\rm Tr}\rho_i^2)$, 
 which becomes 
  \begin{equation}
 S_2(\rho^{\pm}_i)=4p^{\pm}_+p^{\pm}_-=\frac{(1-\cos^{4s}\theta_i)(1-\gamma_i^2)}{(1\pm\langle -\Theta|\Theta\rangle)^2}\label{S2}\,.
 \end{equation}
 For $s=1/2$, the entropy \eqref{S2} and the magnetization \eqref{Sz} are directly related: Eq.\ (\ref{rhoi}) becomes diagonal in the standard basis  $\{|0\rangle=|\!\uparrow_i\rangle,|1\rangle=|\!\downarrow_i\rangle\}$, i.e.\ 
 $\rho_i^{\pm}=p_+^\pm
 |0\rangle\langle 0||+p_-^\pm
 |1\rangle\langle 1|$, 
 and hence $\langle S_i^z\rangle_{\pm}=\frac{p_+^{\pm}-p_-^{\pm}}{2}$, implying  
 \begin{equation}S_2(\rho_i^{\pm})=1-4\langle S_i^z\rangle^2_{\pm}\;\;\;\;(s=1/2)\,.\label{sum}\end{equation}
 Thus, zero local magnetization corresponds in this case to maximum spin-rest entanglement. 
 
 On the other hand, the reduced state  $\rho_{ij}={\rm Tr}_{k\neq ij}|\Psi^{\pm}_-\rangle\langle\Psi^{\pm}_-|$ of two spins $i\neq j$ is 
 \begin{equation}
     \rho_{ij}^{\pm}=
     {\textstyle\frac{|\Theta_{ij}\rangle\langle \Theta_{ij}|+|-\Theta_{ij}\rangle\langle-\Theta_{ij}|\pm\gamma_{ij}(|\Theta_{ij}\rangle\langle-\Theta_{ij}|+|-\Theta_{ij}\rangle\langle\Theta_{ij}|)}{2(1\pm\langle-\Theta|\Theta\rangle)}}\label{rij}
 \end{equation}
 where $|\Theta_{ij}\rangle=|\theta_i\theta_j\rangle$ and 
 $\gamma_{ij}=\prod_{k\neq i,j}\cos^{2s}\theta_k$. 
 It is again a rank $2$ mixed state with eigenvalues  similar to (\ref{ppm}) ($\gamma_i\rightarrow \gamma_{ij}$, $\cos^{2s}\theta_i\rightarrow \cos^{2s}\theta_i\cos^{2s}\theta_j$).  Its quadratic entropy, 
 measuring the entanglement of the pair with the rest of the chain, is then given by an expression similar to (\ref{S2}). 
 Analogous  expressions hold for reduced states of any group of spins \cite{RCM.08}.
 
 A remarkable property of the pair state (\ref{rij}) is that it  depends on the angles $\theta_i$ and $\theta_j$  but not on the actual distance between the spins.  Hence, 
 the entanglement between spins $i$ and $j$ 
 in the state \eqref{rij}, though weak (but finite in a small chain), will be independent of the spin separation for fixed angles $\theta_{i(j)}$. Such entanglement measures the deviation of $\rho_{ij}$ from a separable mixed state, i.e., from a convex mixture of product states.  
 
 Since $\rho_{ij}$  is a rank $2$ mixed state with rank $2$ reduced states, it can be viewed as an effective two-qubit system and the pair entanglement can be quantified 
 through the corresponding concurrence \cite{RCM.08,W.98}. The result is 
 \begin{equation}
     C(\rho_{ij}^{\pm})={\textstyle\frac{\gamma_{ij}\sqrt{(1-\cos^{4s}\theta_i)(1-\cos^{4s}\theta_j)}}{1\pm\langle-\Theta|\Theta\rangle}\label{Cij}}\,,
 \end{equation}
 which  is of parallel (antiparallel) type \cite{Am.06} for positive (negative) parity, with $C(\rho_{ij}^-)\geq C(\rho_{ij}^+)$. It is thus verified to be independent of the separation, being determined just by the angles $\theta_i,\theta_j$ and the complementary overlap $\gamma_{ij}$.  For an alternating state $|\Theta\rangle$, just three concurrences are then obtained at factorization:  $C_{11}$ and  $C_{22}$ for $\theta_i=\theta_j=\theta_{1}$ or $\theta_2$ and $C_{12}$ for $\theta_i=\theta_1$, $\theta_j=\theta_{2}$.    
 Pairwise entanglement then reaches {\it full range} at the factorizing curve, although it becomes rapidly small as size increases due to the factor $\gamma_{ij}$,  in agreement with monogamy \cite{CKW.00}. 
 
 We finally note that if the whole system reduces to a  single spin-$s$ pair,  Eqs.\ \eqref{Sz} and \eqref{Cij} become 
 \begin{eqnarray}
 \langle S_i^z\rangle_{\pm}&=&
 s\frac{\cos\theta_i\pm\cos^{2s}\theta_j\cos^{2s-1}\theta_i}{1\pm\cos^{2s}\theta_i\cos^{2s}\theta_j}\label{M22}\,,\\
  C(\rho^{\pm}_{ij})&=&\frac{\sqrt{(1-\cos^{4s}\theta_i)(1-\cos^{4s}\theta_j)}}{1\pm
 \cos^{2s}\theta_i\cos^{2s}\theta_j}=\sqrt{S_2(\rho_i^{\pm})}\,,\;\;\;\;\;\label{S22}
  \end{eqnarray}
  with $\rho_i^{\pm}$ and $\rho_j^{\pm}$  obviously isospectral since $\rho_{12}^{\pm}$ is pure. In this case the concurrence (\ref{Cij}) reduces to the square root of the linear entropy \eqref{S2}, in agreement with the general result for pure two-qubit states \cite{W.98}. For $s=1/2$ Eq.\ \eqref{sum} is again verified. These expression can be directly expressed in terms of the factorizing fields and coupling strengths through Eq.\ \eqref{thij}. 

  \subsection{Results}
We now show results for the GS magnetization and entanglement in some selected spin pairs and chains, in order to visualize the role  of the factorizing transition. 

 \begin{figure}[ht]
\vspace*{.5cm}	
\centerline{\scalebox{0.06}{\includegraphics
{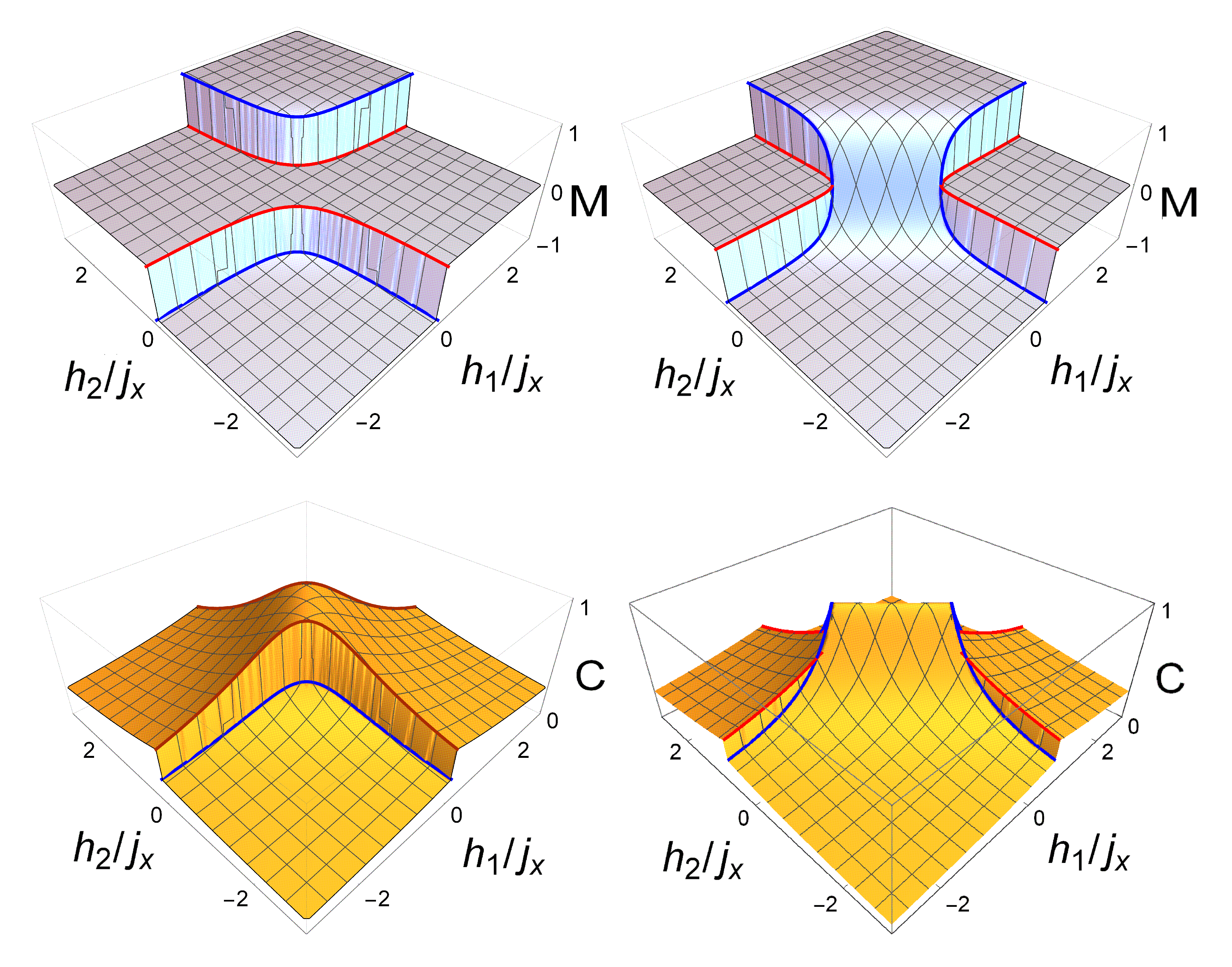}}}
\caption{Ground state magnetization  $M=\langle S_1^z+S_2^z\rangle$ (top) and entanglement (bottom), measured through the concurrence $C$,  of  a  spin $1/2$ pair as a function of the scaled magnetic fields $h_i/j_x$ at each spin. The $XYZ$ couplings 
 are $j_y=0.5 j_x$ and   $j_z=0.25 j_x\, (0.75 j_x)$ on the left (right) panels. Solid  lines indicate the side limits at  the factorizing curves, which determine the GS parity transitions (both $C$ and $M$ are dimensionless).}
\label{f5}
\end{figure}
 
Fig.\ \ref{f5} depicts the total GS magnetization 
$M=\langle S_1^z+S_2^z\rangle$ and  concurrence $C(\rho_{12})$ of a spin $1/2$ pair. The 
negative parity sectors coincide in this case exactly with the zero magnetization plateau,  as is apparent from Eqs.\ \eqref{psipm} and also \eqref{M22} ($\langle S_1^z\rangle_-=-\langle S_2^z\rangle_-$).  For 
$j_z<j_y$ (top left), we see that the positive  parity sectors are also associated with approximate magnetization plateaus, with the factorization curves coinciding with their borders. On the other hand, for $j_z>j_y$ (top right) the magnetization in the positive parity sector evolves continuously from maximum ($1$) to minimum ($-1$). 

The exact concurrence $C(\rho_{12})$ is  in this case  $|\sin 2\gamma^{\pm}_{-}|$, where $\gamma^\pm_-$ are the angles  in the states \eqref{psipm}. It  is larger for negative parity when $j_z<j_y$ (bottom left panel), saturating in this sector for $h_1=h_2$,  where $C(\rho_{12})=1$ ($|\gamma^-_-|=\pi/4$). In contrast, for $j_z>j_y$ the maximum value is attained along the $h_1=-h_2$ line in the positive parity sector,  where again $C(\rho_{12})=1$.   
Note that for $s=1/2$ Eqs.\ \eqref{M22}--\eqref{S22} lead to the values (side-limits)
\begin{eqnarray}
M_{\pm}&=&{\left\{\begin{array}{c}
\frac{\cos\theta_1+\cos\theta_2}{1+\cos\theta_1\cos\theta_2}\\0\end{array}\right.}\,,\;\;\;
C_{\pm}=\frac{|\sin\theta_1\sin\theta_2|}{1\pm\cos\theta_1\cos\theta_2}\nonumber\,,
\end{eqnarray}
at the factorization curves, with 
$\theta_{1(2)}$ determined by (\ref{thij}). 
For $j_z<j_y$, it is then verified that 
$C_{-}=1$ when $h_1=h_2$  ($\theta_1=\theta_2$) and $C_+=1$ when $h_1=-h_2$ ($\theta_1=\pi-\theta_2$).  

\begin{figure}[ht]
\vspace*{.5cm}
\centerline{\scalebox{0.06}{\includegraphics
		{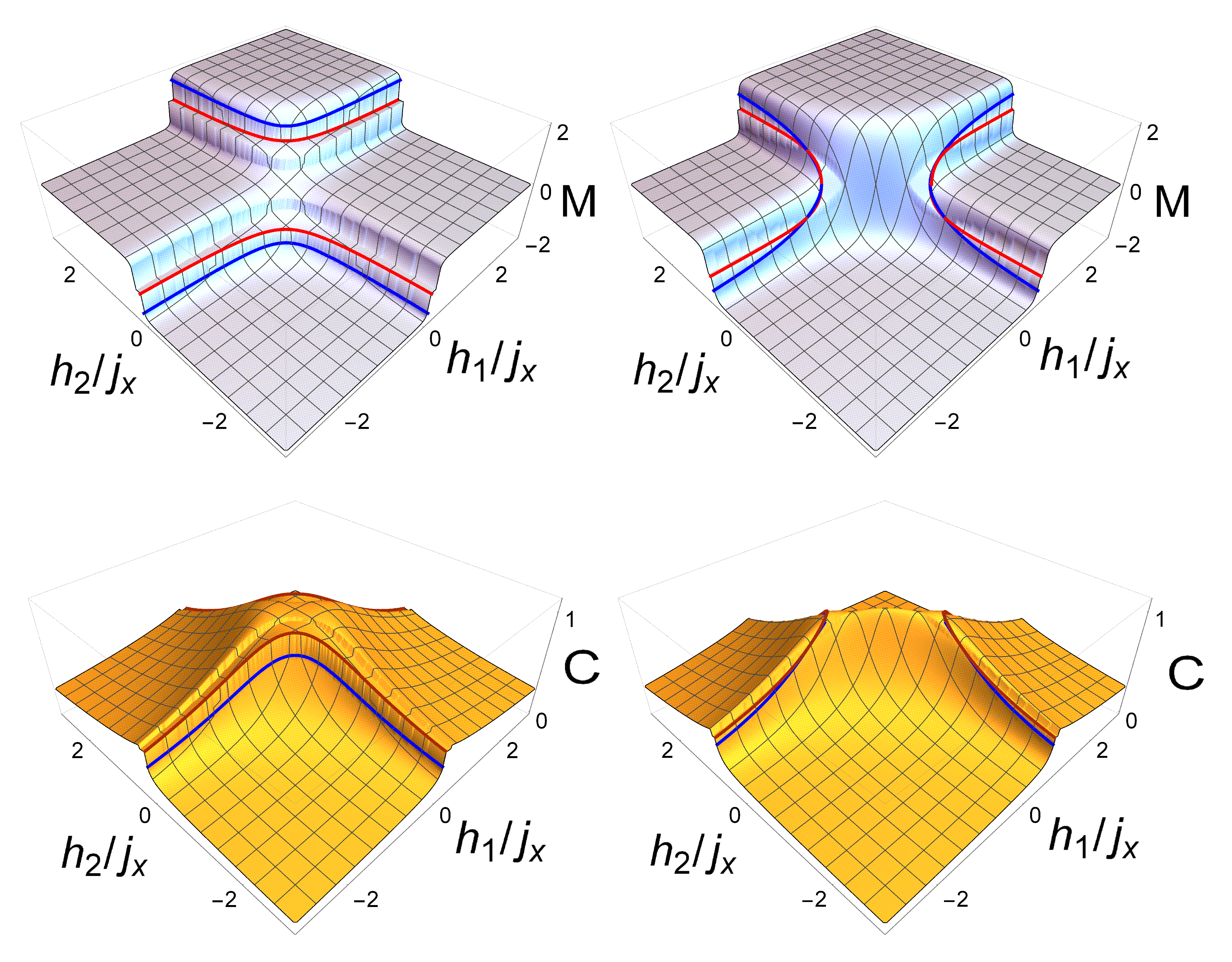}}}
\caption{Ground state magnetization (top) and entanglement (bottom), measured through $C=\sqrt{S_2(\rho_i)}$,   for a  spin $1$ pair as a function of the scaled magnetic fields $h_i/j_x$ for $j_y=0.5 j_x$  and $j_z=0.25 j_x$ (left) and $0.75 j_x$ (right).  Details are similar to those of Fig. \ref{f5}.}
\label{f6}
\end{figure}

Results for the spin $1$ pair are shown in Fig.\ \ref{f6}. In agreement with the parity  diagrams of the left panels in Fig.\ \ref{f3}, the plots show now four steps and five approximate plateaus,  with the factorization curves determining one of the steps (the last one for $j_z<j_y$ when viewed from the origin). The discontinuities at the factorization curve are now  smaller due to the decreased overlap $\langle -\Theta|\Theta\rangle$, and  the (approximate) zero magnetization plateau ($M$ is now not strictly constant in any sector) corresponds to the first even parity sector. 
Results are otherwise similar to the previous case. We have measured the pair entanglement through the square root of the linear entropy, $C=\sqrt{S_2(\rho_i)}$, such that  the values at the border of the factorization are given by  Eq.\ (\ref{S22}). 

\begin{figure}[ht]

\centerline{\scalebox{0.06}{\includegraphics
		{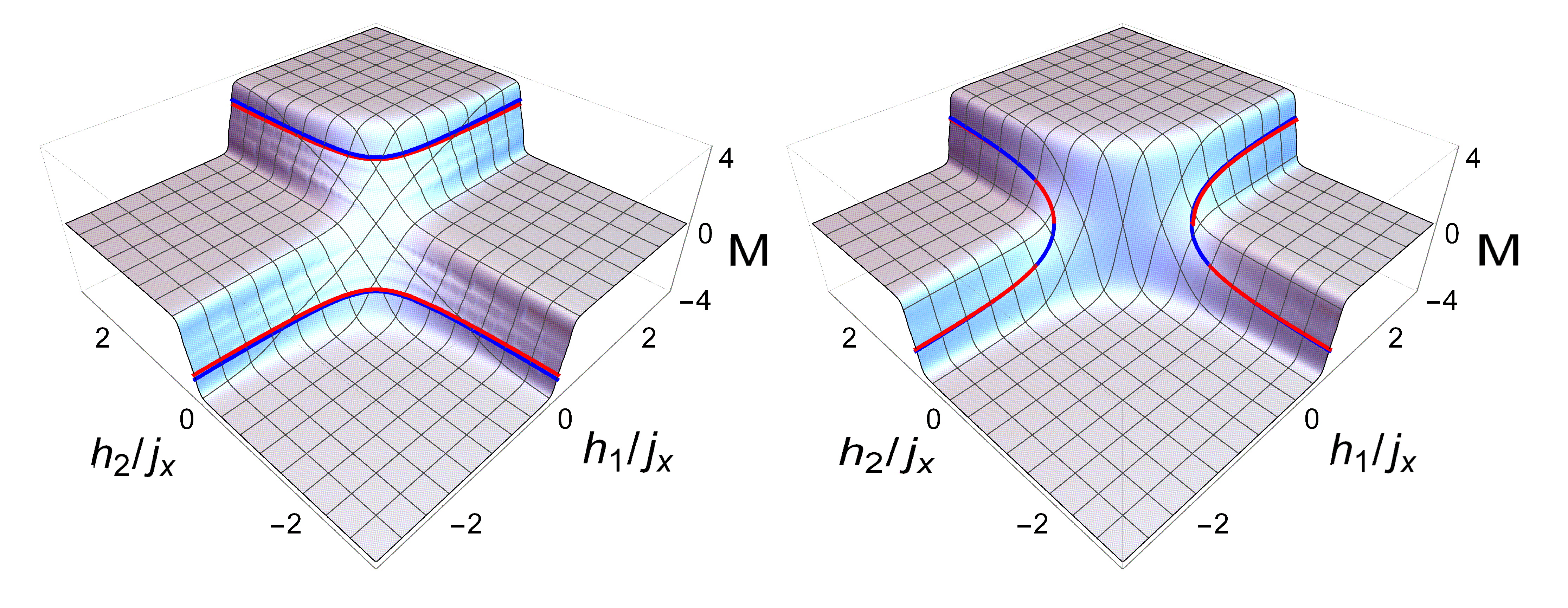}}}
\caption{Ground state magnetization  $M=\langle \sum_i S_i^z\rangle$ of a  spin $1/2$ cyclic chain of 8 spins and  $XYZ$ Heisenberg couplings with  $j_y=0.5 j_x$ and $j_z=0.25 j_x$  $(0.75 j_x)$ on the  left (right) panels, as a function of the scaled magnetic fields $h_i/j_x$ at each site. Solid  lines depict the magnetization at the  factorization curves.}
\label{f7}
\end{figure}

\begin{figure}[ht]
\centerline{\scalebox{.095}{\includegraphics
		{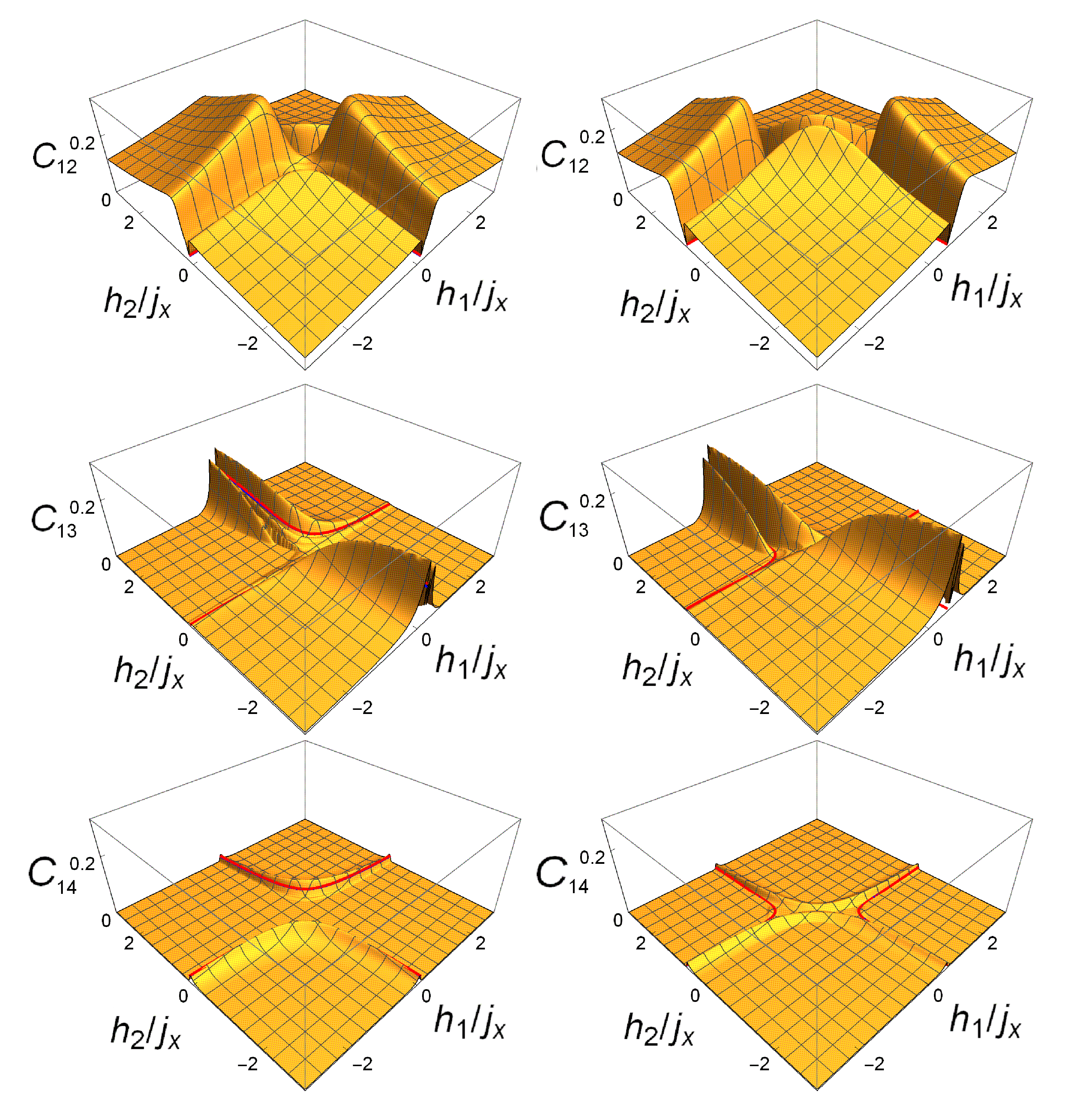}}}
\caption{Ground state pairwise concurrence for first (top), second (center) and third  (bottom) neighbors for $j_y=0.5 j_x$ and $j_z=0.25 j_x$ (left) and $0.75 j_x$ (right), for  the  same chain of Fig.\ \ref{f7},  as a function of the scaled magnetic fields $h_i/j_x$. Solid  lines depict the side-limits at the factorizing fields.}
\label{f8}
\end{figure}

Finally, Figs.\ \ref{f7}--\ref{f8} depict illustrative results for a cyclic spin $1/2$ chain of $n=8$ spins in an alternating field. The magnetization plots (Fig.\ \ref{f7}) are similar to those of a spin pair, but  the magnetization steps at the parity transitions (indicated in the right panels of  Fig.\ \ref{f3})  are now very small, including that at the factorization curves: Results from the states $|\Psi^{\pm}_-\rangle$ (Eq.\ \eqref{Sz}, blue and red solid lines) are very close   due to the small overlap $\langle -\Theta|\Theta\rangle$.  

The corresponding concurrences of first,  second  and third neighbors are depicted in Fig.\ \ref{f8}. Now the values at factorization, determined by Eq.\ \eqref{Cij}, become very small due to the overlap factor $\gamma_{ij}$.  
In the case of first neighbors, the factorization curves appear then  as deep valleys (top panels), since their  concurrence $C_{12}$ is significant away from factorization. In this case maximum concurrence is attained for finite  opposite fields in both cases. In contrast, for third neighbors (bottom), the concurrence $C_{14}$ is nonzero mainly in the vicinity of the factorizing curve for $j_z<j_y$ (left) and close to the outer (non-factorizing) GS parity transition 
for $j_z>j_y$ (right). Note also that at the border of  factorization, 
this concurrence is the same as that for first neighbors ($C_{14}=C_{12}$  according to Eq.\ \eqref{Cij}).  On the other hand, for second neighbors (center), the concurrence $C_{13}$ is maximum when the corresponding field $h_1$ is small, even increasing  
when the intermediate field $h_2$ becomes large, 
 since first neighbors become weakly  entangled due to the alignment of one of the spins.  Nonetheless, small but finite values are still observed in the vicinity of the factorizing  curve. 
 \vspace*{-0.5cm}
 
\section{Conclusions\label{IV}}
We have analyzed GS factorization in finite spin-$s$ arrays with anisotropic XYZ couplings under nonuniform transverse magnetic fields. We have shown that it is essentially a spin-independent phenomenon arising at a fundamental GS parity transition present for any
spin and size, where the GS becomes two-fold degenerate and a
pair of parity breaking product GS’s become exactly feasible. Starting with the case of a spin pair, the general equation
\eqref{eq} for the factorizing fields was derived, 
together with simple analytic expressions for the GS energy and the parameters of the factorized GS. These results generalize those obtained for uniform fields or more symmetric cases and directly
imply the existence of alternating product GS’s in spin chains and square-type arrays with first neighbor XYZ couplings under essentially alternating factorizing fields (with border corrections in open cases), which satisfy the same equation \eqref{eq} when adequately scaled. Moreover, 
alternating fields can induce GS factorization for all values of the couplings in both ferromagnetic and antiferromagnetic systems, including those cases  where no factorization under uniform fields arises. 

 We have also determined the GS parity diagram in field space. 
 It shows a cascade of $2\times ns$ parity transition curves in all cases, and exhibits two distinct regimes according to the $yz$ anisotropy: one for systems with uniform factorizing field (type a)  diagrams, top panels in Fig.\ \ref{f3}), where factorization corresponds to the ``last'' parity transition as seen from the origin, the other for systems without (type b) diagrams, bottom panels), where the pattern becomes more complex and involves factorizing fields of opposite sign only. The  $XZZ$ system (central panels) represents an intermediate critical case  where all GS parity transition curves intersect at zero field. 

Related aspects like  entanglement and magnetization and their behavior in the vicinity of factorization were also analyzed. The factorization curves  represent an entanglement transition and lead to critical entanglement properties in their vicinity. Analytic expressions for these limits in this general setting were provided. 

In summary, the present results unveil new features of the factorization phenomenon in finite $XYZ$ systems under nonuniform fields and their relation with parity symmetry. 
Factorization enables the knowledge of the exact GS of these strongly correlated systems  at least along  certain curves in field space, allowing  insights into the magnetic properties and the complex behavior of quantum correlations. It also enables to cool down an exactly separable non degenerate GS in a strongly interacting system by application of a suitable field, which 
can be useful for quantum simulation \cite{N.14}, quantum  annealing \cite{DC.08} and quantum protocols \cite{VC.05}  based on a fully separable  initial state. The increasing possibilities of simulating spin systems with tunable couplings and fields through different platforms \cite{S.11,L.12,N.14,Guy.18,N.14,PC.04,KK.09, B.12,A.16}  could provide a useful mean for testing and extending the present results.

\acknowledgments
Authors acknowledge support from CONICET (NC) and  CIC (RR)  of Argentina, and from R\'egion Auvergne-Rh\^one-Alpes and
 ENS de Lyon, France (RM).   
 Work supported by CONICET PIP Grant 112201501-00732.
\appendix*
\section{The spin-$s$ pair \label{c}} 
For general spin  $s$ and $j_y\neq j_x$,  we show here that when 
 the fields satisfy Eq.\ (\ref{eq}), a GS parity crossing always takes 
place, at which 
the GS becomes twofold degenerate and a pair of parity breaking product states $|\pm\Theta\rangle=|\pm\theta_1,\pm\theta_2\rangle$ 
become  GS's. \\
{\it Proof:} We choose the $x$ axis such that  $|j_x|>|j_y|$ and set $j_x>0$, as the case $j_x<0$ is related with the former just by a local rotation at one of the spins. For $j_x>0$, minimum $\langle  H\rangle_{\Theta}$ (GS factorization) requires $\theta_1,\theta_2$ of {\it the same sign} (in the interval $(-\pi,\pi)$), as seen from  (\ref{ee1}).   

For fields satisfying  Eq.\ (\ref{eq}) we then choose positive angles $\theta_{1(2)}\in(0,\pi)$ fulfilling  (\ref{thij}) and Eqs.\  (\ref{eqq}). The condition $|j_y|<j_x$ and 
Eq.\ (\ref{eq}) ensure that the quotient in (\ref{thij}) is nonnegative and $<1$. In such a case, $|\Theta\rangle=|\theta_1,\theta_2\rangle$ is an exact eigenstate of $H$, with  $|\theta_i\rangle=R_i|\!\uparrow_i\rangle$ given for general spin $s$  by 
\begin{equation}
|\theta_i\rangle=\sum_{m=-s}^s\!\!\sqrt{\left(^{\;\;2s}_{s-m}\right)}\cos^{s+m}\tfrac{\theta_i}{2}\sin^{s-m}\tfrac{\theta_i}{2}|m_i\rangle\,.\label{thi}
\end{equation}
Hence, the expansion coefficients of 
$|\Theta\rangle$ in the standard product basis $\{|m_im_j\rangle\}$ ($S_i^z|m_i\rangle=m_i|m_i\rangle$) are all non-zero and of the same sign. Since the non-zero off-diagonal elements of $H$ in the previous basis  
arise from 
\[-\sum_{\mu=x,y}\!\!j^{ij}_\mu S_i^\mu S_j^\mu=-j_+(S_i^+S_j^{-}+S_i^-S_j^+)-j_-(S_i^+S_j^++S_i^-S_j^-)\] 
where $S_i^{\pm}=S_i^x\pm iS_i^y$ and $j_{\pm}=(j_x\pm j_y)/4$,  
they are all negative if $|j_y|<j_x$. Hence, a GS $|\Psi\rangle=\sum_{m_1,m_2}C_{m_1 m_2}|m_1m_2\rangle$  with  $C_{m_1 m_2}\geq 0$ $\forall$ $m_1,m_2$ exists, as different signs will not decrease  $\langle \Psi|H|\Psi\rangle$.    
Therefore,  $|\Theta\rangle$  {\it must be a ground state} 
since it cannot be orthogonal to $|\Psi\rangle$. The same holds for $|-\Theta\rangle=P_z|\Theta\rangle$  since $[H,P_z]=0$, implying GS degeneracy.  For such angles (and $|j_y|< j_x$), $\varepsilon_{\Theta}\pm j_z<0$ in (\ref{aux}) since Eq.\ (\ref{eq}) was originally fulfilled,  implying that   $\varepsilon_\Theta=E_\Theta/s$ 
will be given by Eqs.\ (\ref{ee34}). The connection between the separable states $|\pm\Theta\rangle$ and the crossing definite parity GS's $|\Psi^{\pm}\rangle$ at factorization
 will be given again by (\ref{rel}), i.e.\ 
 $|\Psi^{\pm}\rangle=\frac{|\Theta\rangle\pm|-\Theta\rangle}{2(1\pm\langle-\Theta|\Theta\rangle)}$.  Through the indicated scalings of fields and couplings, these arguments can be directly applied to a pair with distinct spins $s_i\neq s_j$, as well as to alternating solutions in the $XYZ$ spin chains and arrays of section \ref{II}, provided $|j^{ij}_y|<j^{ij}_x$ for all coupled pairs. 

 For small deviations $\delta h^i$ of the fields from the factorization curve, at fixed couplings  we have a  variation $\delta E_{\pm}\approx -\sum_i \delta h^i \langle S^z_i\rangle_\pm$ of the energies of the definite parity GS's,  where $\langle S^z_i\rangle_{\pm}$ are the averages \eqref{Sz} in the  states $|\Psi^{\pm}\rangle$. The difference $\langle S^z_i\rangle_--\langle S^z_i\rangle_+$ (magnetization jump) then leads to the energy splitting 
 \begin{equation}\delta E_--\delta E_+\approx {\textstyle 2s\frac{\langle -\Theta|\Theta\rangle}{1-\langle-\Theta|\Theta\rangle^2}\sum_i \frac{\sin^2\theta_i}{\cos\theta_i}\delta h^i}\,,\end{equation} 
 which is small but non-zero in a finite array for variations off the factorization curve. This shows that the degeneracy of the opposite parity GSs  will be lost for small deviations away from factorization, so that the factorization curve corresponds to a GS parity transition. \qed


\begin{thebibliography}{999}
\bibitem{ft}If $j_y>|j_x|$ we should set $\phi_{1(2)}=\frac{\pi}{2}$, which leads to Eqs.\ \eqref{eq11}--\eqref{eq33} with $j_x\leftrightarrow j_y$. And the case $(\theta,\phi=\pi)$ is obviously equivalent to $(-\theta,0)$.
\bibitem{ft2}If $\sin\theta_1=0$ and $j_x\neq 0$, \eqref{eq11}--\eqref{eq33} imply $\sin\theta_2=0$ and $j_y=j_x$, which is the $XXZ$ case considered in \cite{CRC.17}.
\end{thebibliography}
\end{document}